\documentclass[sigconf]{acmart}
\setcopyright{none}
\copyrightyear{}
\acmYear{}
\acmDOI{}
\acmISBN{}

\acmConference{}
{}{}{}

\renewcommand\footnotetextcopyrightpermission[1]{}

\begin{document}

\title{Discovery-Oriented Faceting: From Coverage to Blind-Spot Discovery}

\author{Youdi Li}
\email{ri.yutei@jp.panasonic.com}
\affiliation{%
  \institution{Panasonic Connect Co., Ltd.}
  \city{Tokyo}
  \country{Japan}
}

\renewcommand{\shortauthors}{Li}

\begin{abstract}
 When people explore large document collections to build understanding, they face a challenge: existing AI tools help them see what is central but tend to hide what is unusual. Summarization and topic modeling optimize for coverage, representing main themes while pushing minority viewpoints and edge cases out of view. This matters because discovery often depends on noticing what does not fit, such as unexpected findings, minority positions, or gaps in the literature. When tools hide this content, users may miss insights that could change their understanding. In this paper, we explore an alternative objective: blind-spot discovery, where the goal is to surface content that coverage methods suppress so that people can judge its significance for themselves. We propose three design goals and illustrate them through DOF (Discovery-Oriented Faceting), a system that organizes documents into categories with explicit boundaries, ranks categories by distinctiveness rather than size, and supports iterative refinement. Comparing DOF against coverage-based ranking across four domains, we find that the two approaches surface fundamentally different content, with DOF promoting specialized categories that coverage methods bury. We discuss how shifting from coverage to discovery may offer a complementary mode of support for people exploring large text collections.

\end{abstract}

\maketitle
\pagestyle{plain}

\section{Introduction}

Making sense of large document collections is central to knowledge work. Patent examiners search archives to spot unexpected approaches. Literature researchers survey prior work to find gaps. Policy analysts review reports to identify emerging issues.

AI tools for document exploration, including faceted summarization~\cite{meng2021facetsum}, topic modeling~\cite{blei2003latent, grootendorst2022bertopic}, and interactive navigation~\cite{hirsch2021ifacetsum}, help structure large text collections. However, current tools share a design choice that works against discovery: they optimize for coverage. FacetSum~\cite{meng2021facetsum} and WikiAsp~\cite{hayashi2021wikiasp} use predefined categories that represent central content well but miss minority viewpoints. Topic models produce clusters that overlap and drift, with boundaries hard to interpret~\cite{smith2018interactive}. LLM-guided approaches~\cite{zhang2023clusterllm} improve cluster coherence but do not expose distinctive content. Once generated, most structures offer no way to refine~\cite{hirsch2021ifacetsum}.

The problem is that coverage optimization pushes unusual content out of view. The content most valuable for new understanding (specialized vocabulary, minority positions, emerging themes) is systematically deprioritized. When tools show only the central tendency, they risk reinforcing what people already know rather than exposing what they do not. Recent work addresses parts of this problem~\cite{delort2012dualsum, conroy2011nouveau, iso2022cocosum, cohn2003semisupervised}, but each only partly: surfacing novelty without ranking, or accepting feedback without explicit boundaries. Ebrahimi and Peltonen~\cite{ebrahimi2025constrained} used constrained matrix factorization for minority topic discovery, though their method requires predefined seed words.

In this paper, we present Discovery-Oriented Faceting (DOF), a system that takes an alternative approach: blind-spot discovery, where the goal is to reveal content that standard methods suppress. DOF is built around three design goals:

\textbf{D1. Make boundaries explicit.} Each category specifies what belongs and what does not. A topic model might label a cluster ``Drug Research,'' leaving users unsure whether it includes clinical trials or pricing policy. DOF specifies inclusions and exclusions so users can quickly judge relevance.

\textbf{D2. Prioritize what is unusual.} DOF ranks categories by how much they differ from the corpus average, not by size, highlighting content that is both coherent and distinctive.

\textbf{D3. Support iterative refinement.} Understanding develops through exploration~\cite{pirolli2005sensemaking}, and DOF is designed to intervene at the transition from information foraging to schema construction (see Section~\ref{sec:discussion}). DOF lets users merge, split, or hide categories without starting over.

We evaluate DOF across four domains, showing that distinctiveness ranking reveals specialized content that coverage methods miss. We discuss implications for tools that support discovery in knowledge work.

\section{Related Work}

\subsection{Faceted Summarization}
Latent Dirichlet Allocation~\cite{blei2003latent} established probabilistic topic modeling, with BERTopic~\cite{grootendorst2022bertopic} extending this via neural embeddings. FacetSum~\cite{meng2021facetsum} introduced structured summaries with predefined facets, and WikiAsp~\cite{hayashi2021wikiasp} extended this to 20 domains. iFacetSum~\cite{hirsch2021ifacetsum} enabled interactive navigation via induced facets. Boytsov et al.~\cite{boytsov2025aspectguided} extracted aspect-sentiment pairs from reviews, though specialized to sentiment-bearing domains. These systems relied on static schemas or domain-specific pipelines, leaving open how to discover facets dynamically.

\subsection{Novelty-Aware Summarization}
DualSum~\cite{delort2012dualsum} separated novel from redundant content, and Nouveau-ROUGE~\cite{conroy2011nouveau} proposed metrics rewarding absent content. CoCoSum~\cite{iso2022cocosum} generated contrastive summaries between entity sets. STRUM~\cite{gunel2023strum} and Hayashi et al.~\cite{hayashi2023disentangled} generated ``contribution'' vs.\ ``context'' summaries without predefined ontologies. Ebrahimi and Peltonen~\cite{ebrahimi2025constrained} incorporated seed word constraints into NMF to capture low-prevalence topics, though users must specify the seed words in advance. These approaches treated novelty as binary rather than as a ranking criterion for exploration.

\subsection{Human-in-the-Loop Refinement}
Interactive clustering with user constraints~\cite{cohn2003semisupervised} improved coherence, while Smith et al.~\cite{smith2018interactive} found users over-trusted topic models. ClusterLLM~\cite{zhang2023clusterllm} used LLM feedback to improve cluster quality, focusing on coherence rather than distinctiveness. Fang et al.~\cite{fang2023interactive} developed interactive topic modeling with change tracking, and Zade et al.~\cite{zade2018disagreement} showed how visualizing disagreement aided interpretation. CollabCoder~\cite{gao2024collabcoder} and Diaz-Rodriguez~\cite{diazrodriguez2025centroids} integrated model suggestions with collaborative coding and proposed ``summaries as centroids.'' However, existing systems lacked refinement mechanisms or required full regeneration after each edit.

\section{The DOF System}

\begin{figure*}[t]
\centering
\includegraphics[width=\textwidth]{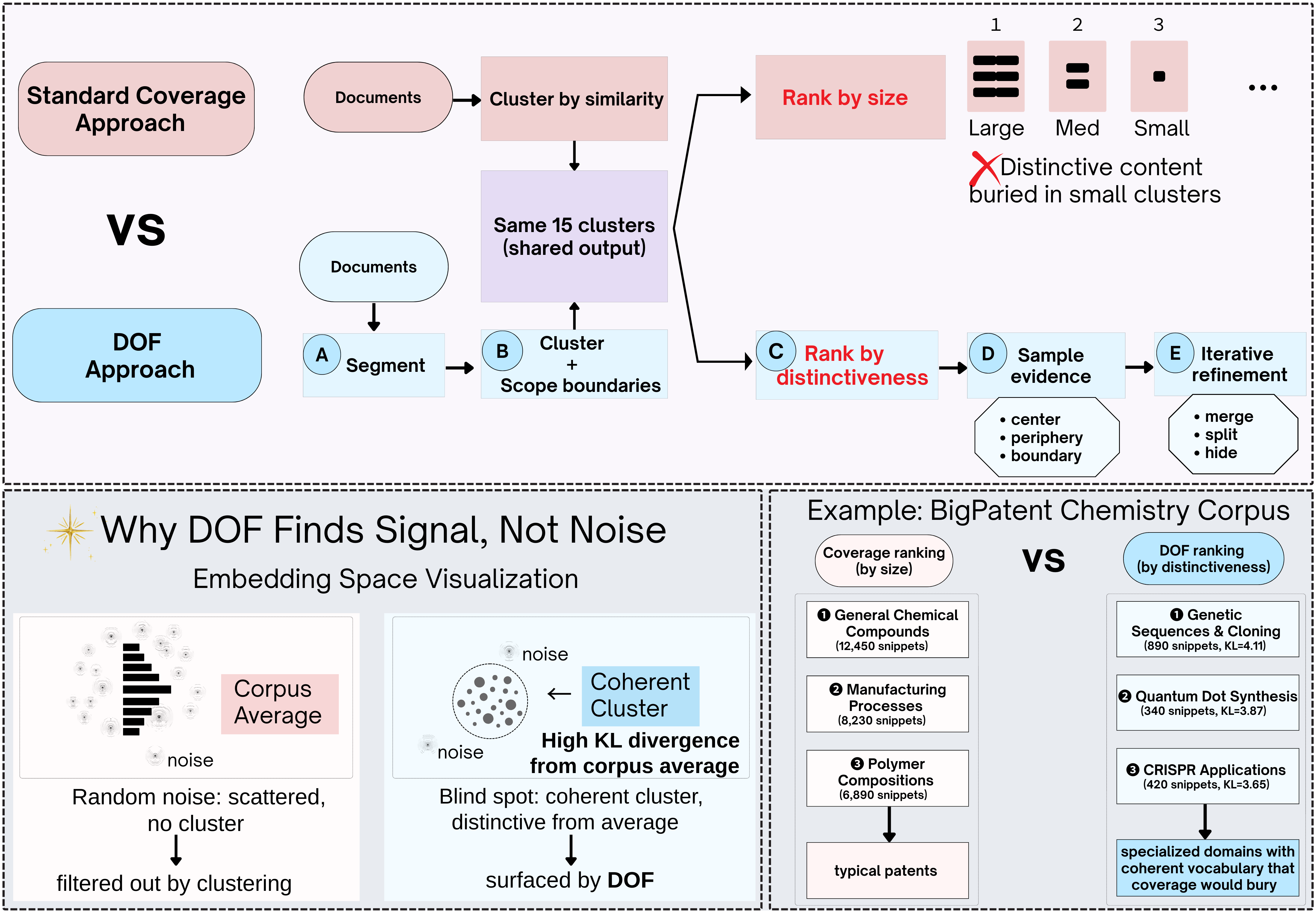}
\caption{DOF contrasted with coverage approaches. \textbf{Upper:} Both approaches cluster documents, but the key difference is ranking. Coverage ranks by size (large clusters first), burying distinctive content. DOF ranks by KL divergence (distinctive first), then adds scope boundaries, evidence sampling, and user refinement. \textbf{Lower left:} Signal vs.\ noise in embedding space. Random noise scatters without forming clusters, while coherent blind spots form distinct clusters with high KL divergence, which DOF prioritizes. \textbf{Lower right:} Illustrative BigPatent example showing the same clusters ranked differently. Coverage foregrounds large, typical clusters while DOF foregrounds small, specialized ones.}
\Description{A three-part diagram. Upper panel shows a flowchart comparing coverage and DOF pipelines, with shared clustering step followed by divergent ranking strategies. Lower left shows embedding space with scattered noise points versus coherent blind-spot clusters. Lower right shows ranked lists comparing coverage (size-based) versus DOF (distinctiveness-based) outputs on BigPatent data.}
\label{fig:pipeline}
\end{figure*}

Figure~\ref{fig:pipeline} illustrates the DOF pipeline and contrasts it with coverage-oriented approaches. The key distinction occurs \emph{after} clustering: coverage tools rank clusters by size (surfacing typical content), while DOF ranks by distinctiveness (surfacing content that deviates from corpus norms). The pipeline has five stages, with all parameters locked before processing and held constant across all datasets.

\subsection{Snippet Construction}
Long documents are segmented into 425-word snippets with 75-word overlap. Each snippet is embedded using OpenAI's embedding model (text-embedding-3-small, 1536 dimensions)~\cite{openai2024embedding}. This fixed granularity avoids the over-generalization that direct LLM application to long documents produces~\cite{liu2024selenite}.

\subsection{Scope-Bounded Clustering}
Snippets are partitioned into $K$=15 clusters via K-Means (random state=42). For each cluster, an LLM (GPT-4o-2024-08-06, temperature=0.7, max tokens=500) examines the snippets and generates a scope definition: a label, four positive statements (what belongs), and four negative statements (what does not belong). The LLM infers exclusions by identifying semantically related content that falls outside the cluster's actual scope. Unlike topic models, which leave boundaries implicit, DOF explicitly states what does not belong~\cite{wang2025schemex}. This helps users decide whether a category fits their needs.

\subsection{Distinctiveness Ranking}
Standard tools rank by size or centrality, burying specialized content. DOF scores each cluster using Kullback-Leibler (KL) divergence between its embedding distribution and the corpus average, then sorts from most distinctive to least. A cluster about prosthetic devices in a chemistry patent corpus scores high because its embedding distribution diverges from the corpus average.

\subsection{Evidence Sampling}
For each top-ranked category, DOF samples three snippets: a \textbf{central} snippet (closest to centroid), a \textbf{peripheral} snippet (farthest valid member), and a \textbf{transitional} snippet (median distance). Showing the range from core to edge lets users verify coherence before exploring further.

\subsection{Iterative Refinement}
Beyond ranking, DOF lets users reshape the structure incrementally: \textbf{merge} categories that overlap, \textbf{split} categories that conflate distinct ideas, or \textbf{hide} irrelevant categories. Each action updates the view in place~\cite{pirolli2005sensemaking}, so refinements build on each other.

\section{Evaluation}

We evaluate DOF on \emph{blind-spot discovery}: whether the system exposes content that deviates from corpus-wide patterns, making it available for human judgment rather than hiding it behind coverage rankings.

\subsection{Datasets}

We selected four datasets spanning distinct domains (Table~\ref{tab:datasets}): \textbf{GovReport}~\cite{huang2021govreport} (U.S.\ government agency reports), \textbf{BillSum}~\cite{kornilova2019billsum} (congressional legislation), \textbf{BigPatent}~\cite{sharma2019bigpatent} (patent applications), and \textbf{BookSum}~\cite{kryscinski2021booksum} (literary works). All were processed with identical locked parameters ($K$=15, 425-word snippets, 75-word overlap). This diversity allows us to observe how DOF generalizes without domain-specific tuning.

\begin{table}[htbp]
\centering
\caption{Dataset overview.}
\label{tab:datasets}
\begin{tabular}{llrr}
\toprule
\textbf{Dataset} & \textbf{Domain} & \textbf{Docs} & \textbf{Snippets} \\
\midrule
GovReport & Government & 973 & 21,001 \\
BillSum & Legislation & 3,269 & 12,872 \\
BigPatent & Patents & 10,764 & 150,000 \\
BookSum & Literature & 1,431 & 17,321 \\
\midrule
\textbf{Total} & & 16,437 & 201,194 \\
\bottomrule
\end{tabular}
\end{table}

\subsection{Ranking Strategy Comparison}

To show that DOF brings different content into view than coverage-based approaches, we compare two ranking strategies on the same clustering results: (1) \textbf{Coverage}: rank by cluster size; (2) \textbf{DOF}: rank by KL divergence. Top-5 overlap counts how many facets appear in both rankings' top five, where lower overlap means users see different content. Spearman $\rho$ measures whether rankings go in the same direction, and $\rho = -1$ means the largest clusters are the least distinctive, so coverage and DOF show opposite content.

\begin{table}[htbp]
\centering
\caption{Coverage vs.\ DOF ranking comparison. Lower overlap and negative $\rho$ indicate different content surfaced.}
\label{tab:baseline}
\begin{tabular}{lcc}
\toprule
\textbf{Dataset} & \textbf{Top-5 Overlap} & \textbf{Spearman $\rho$} \\
\midrule
GovReport & 2/5 & $-$0.06 \\
BillSum & 2/5 & $+$0.31 \\
BigPatent & 0/5 & $-$1.00 \\
BookSum & 0/5 & $-$1.00 \\
\bottomrule
\end{tabular}
\end{table}

In technical domains (BigPatent, BookSum), the correlation is perfectly negative ($\rho = -1.0$): the largest clusters have the lowest distinctiveness scores. Coverage ranking foregrounds typical content (Innovative Personal Care Products, Advanced Personal Care Devices), while DOF highlights specialized domains (Advanced Personal Hygiene Products, Advanced Prosthetic Devices). In BookSum, DOF reveals facets like ``Strategies of Political Leadership'' (\textit{The Prince}) and ``Colonial Confrontations'' (\textit{Lord Jim}), which coverage methods rank lower in favor of more common literary themes.

Government and legislative text shows weaker divergence because formulaic language means even distinctive facets tend to be moderately sized.

\subsection{Cross-Domain Results}

\begin{table}[htbp]
\centering
\caption{Distinctiveness results across domains.}
\label{tab:distinctiveness}
\begin{tabular}{lccc}
\toprule
\textbf{Dataset} & \textbf{KL$_A$} & \textbf{Max KL} & \textbf{Top Facet} \\
\midrule
GovReport & 0.98 & 1.74 & Trade \\
BillSum & 1.08 & 1.97 & Tax \\
BigPatent & 2.78 & 3.59 & Hygiene \\
BookSum & 2.78 & 3.28 & Politics \\
\bottomrule
\end{tabular}
\end{table}

Facet distinctiveness varies systematically across domains (Table~\ref{tab:distinctiveness}). Government and legislative text shows lower distinctiveness (KL$_A$ $\approx$ 1.0) due to formulaic language, while technical and literary text shows higher distinctiveness (KL$_A$ $\approx$ 2.8) due to specialized vocabulary. In BigPatent, ``Hygiene Products'' (Max KL=3.59) has an embedding distribution far from the corpus average, while in GovReport, ``Trade Policy'' (Max KL=1.74) remains closer, reflecting the shared language of government text.

\subsection{Scope Boundaries Aid Interpretation}

Scope boundaries help users understand what belongs in each facet. For example, a BigPatent facet labeled ``Advanced Prosthetic Devices'' explicitly states what it includes (bionic limb technologies, neural interface control systems, adaptive mobility prosthetics, sensory feedback mechanisms) and what it excludes (textile manufacturing, food preservation, water purification, kitchen utensil design). Without these exclusions, users might confuse this facet with broader biomedical or manufacturing content. In GovReport, a facet labeled ``Medicaid and Medicare Administration'' excludes private insurance, international health systems, and pharmaceutical research, helping analysts quickly confirm it matches their task.

\subsection{Preference Refinement}

We simulated 10 rounds of refinement operations across all four datasets to evaluate whether the merge, split, and hide operations help users reach a stable organization (Table~\ref{tab:refinement}). In each round, we applied rule-based operations: merging facets with centroid similarity above 0.8, and hiding facets with KL divergence below the dataset median. We track three metrics: \textbf{inter-facet similarity} (cosine similarity between centroids) measures overlap, \textbf{KL$_A$} measures distinctiveness, and \textbf{edits per round} tracks convergence. On average, similarity decreased by 28\%, KL$_A$ increased by 15\%, and edits dropped by 71\%, suggesting refinement reduces redundancy and focuses on distinctive content.

\begin{table}[htbp]
\centering
\caption{Simulated preference refinement (10 rounds per dataset).}
\label{tab:refinement}
\begin{tabular}{lrrr}
\toprule
\textbf{Dataset} & \textbf{Similarity $\Delta$} & \textbf{KL$_A$ $\Delta$} & \textbf{Edits $\Delta$} \\
\midrule
GovReport & $-$23\% & $+$12\% & $-$68\% \\
BillSum & $-$27\% & $+$14\% & $-$71\% \\
BigPatent & $-$29\% & $+$16\% & $-$72\% \\
BookSum & $-$31\% & $+$18\% & $-$74\% \\
\midrule
\textbf{Average} & $-$28\% & $+$15\% & $-$71\% \\
\bottomrule
\end{tabular}
\end{table}

\subsection{Parameter Robustness and Scope Quality}

Table~\ref{tab:validation} reports three automated checks on GovReport. The inverse relationship between cluster size and distinctiveness holds across $K$=5 to 30, confirming that DOF's ranking is not an artifact of a single $K$ setting. We define a snippet as boundary-ambiguous if its cosine distance ratio to the nearest versus second-nearest centroid exceeds 0.95 (i.e., it nearly belongs to two clusters). At this threshold, 37.4\% of snippets are ambiguous overall, but distinctive clusters have fewer such cases ($\rho = -0.90$, $p < 0.0001$), suggesting DOF prioritizes content that is more cleanly separable. An LLM coherence check (GPT-4o-mini) rated all 15 scope definitions at 4.7/5 on average, with label clarity and inclusion coherence at ceiling (5.0/5) and keyphrase alignment lowest (4.0/5).

\begin{table}[htbp]
\centering
\small
\caption{Automated validation (GovReport, $K$=15).}
\label{tab:validation}
\begin{tabular}{@{}llr@{}}
\toprule
\textbf{Analysis} & \textbf{Metric} & \textbf{Result} \\
\midrule
K sensitivity & $\rho$ (size vs.\ KL) & $-$.73 to $-$.90 \\
($K$=5--30) & Top-5 overlap, $K{\geq}10$ & $\leq$ 1/5 \\
\midrule
Boundary & Ambiguous snippets & 37.4\% \\
membership & $\rho$ (KL vs.\ boundary\%) & $-$.90\textsuperscript{***} \\
\midrule
Scope quality & Label clarity & 5.0/5 \\
(GPT-4o-mini) & Inclusion coherence & 5.0/5 \\
 & Exclusion usefulness & 4.9/5 \\
 & Keyphrase alignment & 4.0/5 \\
\bottomrule
\multicolumn{3}{@{}l@{}}{\textsuperscript{***}$p < 0.0001$}
\end{tabular}
\end{table}

\section{Discussion and Future Work}\label{sec:discussion}

DOF reframes document exploration around what is unusual rather than what is common. Here we situate DOF within sensemaking theory, discuss its intended users, and outline current limitations.

\paragraph{Where DOF Intervenes in Sensemaking.}
Pirolli and Card's~\cite{pirolli2005sensemaking} sensemaking model describes a loop from information foraging (searching, filtering, extracting) to synthesis (building schemas, drawing conclusions). DOF intervenes at the foraging-to-synthesis transition. During foraging, coverage-based tools narrow attention toward frequently occurring themes, potentially causing users to construct schemas that reflect only central content. DOF expands the foraging stage by exposing unfamiliar yet coherent clusters before the user commits to a schema. DOF makes unfamiliar clusters visible and interpretable, but whether users act on this exposure, and how it shapes their final schemas, requires user studies to establish.

\paragraph{Target Users and Use Contexts.}
DOF's value likely varies across use contexts. We see the strongest fit for \emph{exploratory} professionals who face open-ended questions over large corpora: a patent examiner searching for prior art in an unfamiliar technology area benefits from seeing specialized clusters (e.g., prosthetic innovations in a chemistry corpus) that a size-ranked view would bury. A literature researcher surveying a new field may use DOF to identify minority positions that challenge mainstream narratives. Users with well-defined search goals may find distinctiveness ranking less useful or even disorienting, since DOF deliberately foregrounds content outside the familiar center. In such cases, coverage-based tools remain more appropriate. DOF is designed to complement, not replace, them.

\paragraph{Limitations and Future Work.}
The preference refinement used rule-based simulations rather than real user studies. The fixed parameters ($K$=15, 425-word snippets) are robust across $K$=5 to 30 (Table~\ref{tab:validation}), but K-Means with cosine similarity can produce clusters that are mathematically coherent but semantically coarse; LLM scope generation partially compensates but does not guarantee human-meaningful boundaries. User studies and soft clustering are natural next steps.


\bibliographystyle{ACM-Reference-Format}
\bibliography{references}

\end{document}